# Proposal for a dual spin filter based on [VO(C$_3$S$_4$O)$_2$]$^{2-}$

*Salvador Cardona-Serra[1,2*], Alejandro Gaita-Ariño[2], Efrén Navarro-Moratalla[2], Stefano Sanvito[1].*

[1] School of Physics, AMBER and CRANN Institute, Trinity College, Dublin 2, Ireland.

[2] Instituto de Ciencia Molecular, Universitat de València, España.

AUTHOR INFORMATION

**Corresponding Author**

E-mail: salvador.cardona@uv.es

Abstract

Polynuclear magnetic molecules often present dense transmission spectra with many overlapping conduction spin channels. Single-metal complexes display a sparser density of states, which in the presence of a fixed external magnetic field makes them interesting candidates for spin filtering. Here we perform a DFT study of a family of bis- and tris-dithiolate vanadium complexes sandwiched between Au(111) electrodes and demonstrate that $[VO(C_3S_4O)_2]^{2-}$ can behave as a dual spin filter. This means that an external stimulus can switch between the selective transmission of spin-up and spin-down carriers. By using an electrostatic gate as external stimulus we show that the onset for the spin-up conductance is at a voltage $V_g = -0.51$ V but a small shift to $V_g = -0.63$ V is capable of activating spin-down transport. For both cases, we estimate a large low-bias conductance (approx. 2 $\mu$S at $V_{bias} < 50$ mV) with excellent spin selectivity (> 99.5%). We conclude by commenting on the general molecular requirements for the chemical design of further such spintronics components.

**TOC GRAPHICS**

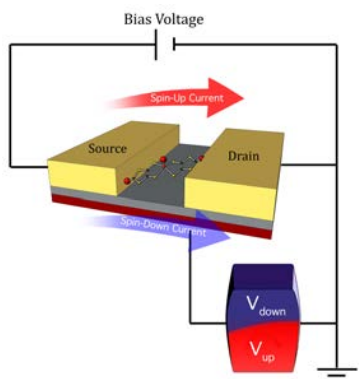

The discovery of the giant magnetoresistance[1] and the tunnel magnetoresistance[2] effects laid down the foundation of spin-based electronics including the possibility of using spins for quantum logic operations. These phenomena are part of the more general concept of spintronics, which relies on the use of the spin instead of the charge to encode information, an approach that permits the manipulation (write/read) of information using lower energies compared to those involved in conventional electronics. In addition in a magnetic material the spin direction remains preserved in the absence of any power[3], namely spintronics devices may be intrinsically non-volatile. Molecular spintronics, developed in the last decade,[4, 5, 6] deals with the possibility of transferring the spintronic properties displayed by the purely inorganic compounds to systems made out of discrete molecules. Motivated by the fact that organic molecules are mostly formed by light atoms, which bear weak spin-orbit coupling and present low contact nuclear hyperfine interaction, molecular spintronics holds promises for enhanced quantum coherence and the preservation of the spin during the operation time; one of the key goals of the scientific community in this field.[7] Recently, the fist quantum algorithm implemented in a single-molecule device by means of spintronics has been reported.[8]

Chemists are playing an important role in this area by experimentally synthesizing a large number of promising molecules, specifically designed for spintronics-based quantum logic. Nowadays, the research focus covers a wide variety of molecules ranging from increasingly large p-conjugated molecules, high mobility (approximately, tens of $cm^2V^{-1}s^{-1}$) systems and hydrogen-free complexes.[4] However, a deeper understanding of the relation between the spintronic behavior of a molecule and its structure is needed in order to address the optimization of the molecular systems *a priori* and help better focus the synthetic efforts. Theoretical calculations are already giving important insights, such as those concerning the mechanism governing the spinterface behavior in Alq$_3$.[9] Additionally, theory can guide the synthetic experimental effort by tailoring and designing molecular systems that can act as functional modules for actual applications. This is the case of rationally designed polar molecules with an spin transition induced by electric field.[10]

Among the many possible applications of spintronics, we focus on dual spin filtering, namely on the possibility of selecting the spin polarization of a device by applying an external stimulus. This is of particular relevance for the implementation of programmable spintronic devices. Given an adequate combination of spin-up and spin-down transport channels, it is in principle possible to switch the sign of the filtered spin within a given molecule-based device.[11] In fact large magnetic molecules containing many magnetic ions typically present multiple conduction channels with either spin-up or spin-down potential [12]. In principle this electronic structure can theoretically implement dual spin filtering, but in practice one requires an unrealistic precision in the control of the gate voltage (the external stimulus). As a result one may expect a poor spin-filtering selectivity. In contrast, small mononuclear complexes may present simpler features, where highly conductive channels are also markedly sharp and can be addressed with precision with a gate voltage. This was for instance the case of some tris-dithiolate complexes.[13] Previous theoretical explorations of spin-filtering effects using single-ion complexes have either assumed markedly unrealistic molecular structures and/or ill-defined external stimuli,[14] or have merely found spin filtering in a single direction, with no possibility for dual filtering. [15, 16] (See SI-4 for a more detailed discussion) In this work we explore the behavior of the family of vanadium bis-and tris-dithiolate, introduced by Freedman et al [17], contacted by two Au(111) electrodes to form a single-molecule dual spin filter device.

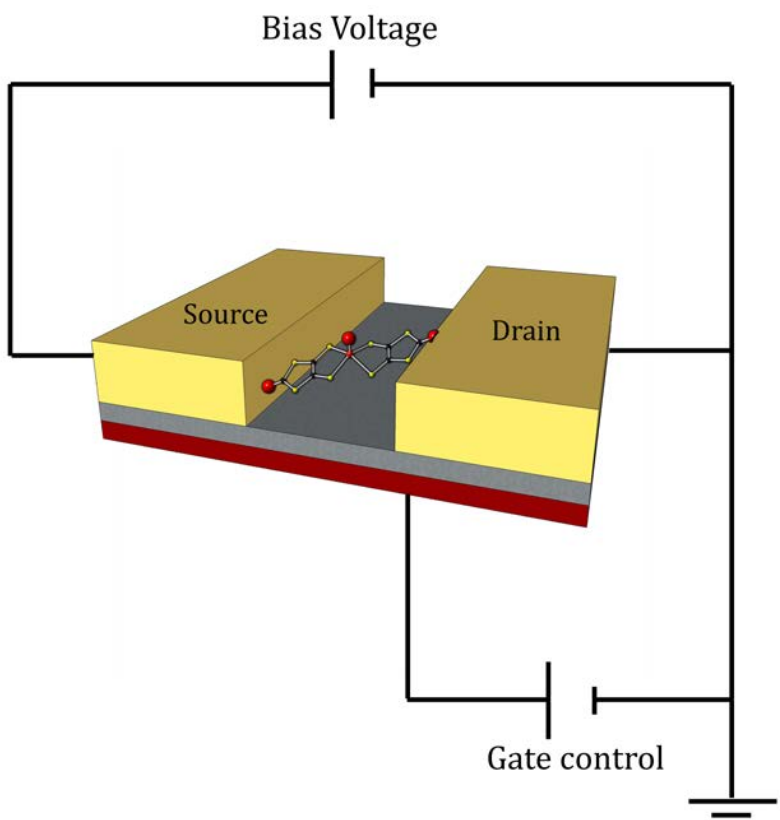

Figure 1: Device scheme of a single-molecule the dual spin filter. A vanadium bis- and tris-dithiolate molecule is contacted to Au (111) electrodes. A backgate is also included in the device setup.

The proposed device structure is presented in Figure 1. The operation of this molecular dual spin filter is based on applying a local electrostatic gate to a single-molecular junction. Here the gate voltage allows the tuning of the chemical potential across the DOS peaks corresponding to up and down spin conductance channels ($E_0=E_{up}-E_{down}/2$). A bias voltage applied to the ends of the molecule drives the current through the junction. This can present different spin polarization depending on the gate voltage, which places the Fermi level in resonance with either a spin up or a spin down channel. Hence, the requirement for this molecular device is that the molecule spectrum has at least two available close-lying but not overlapping spin-up and spin-down conductance channels.

DFT calculations for the entire family of vanadium bis- and tris-dithiolate (using the procedure explained in Ref. 13) display a wide variety of behaviors, but some general trends can be extracted. Firstly, as we known from preliminary calculations,[13] vanadium and vanadyl dithiolate complexes tend to present conduction channels available at the Fermi energy. Certainly, larger complexes, or molecules of higher complexity, tend to display richer spectra. Finally, due to better hybridization, strong S-Au contacts tend to result in wider transmission channels and in single-molecule conductances up to 20 mS at low voltages; steric impediments or weaker O-Au contacts result in sharper channels and conductance of about 2 mS at low voltages.

We consider $[VO(C_3S_4O)_2]^{2-}$ to be the best molecular dual spin filter candidate within the studied family. As can be seen in Figure 2, we find a spin-up conductance at a gate voltage $V_g^{up}$ = -0.51 V, and a small shift to $V_g^{down}$ = -0.63 V switches the system to spin-down conductance. In both cases we calculate an excellent spin selectivity (>99.5%) assuming a full spin polarization of the complex due to a fixed external magnetic field. Concerning the conductances, note that the computational approach we employ does not allow setting a given value for the gate voltage and instead the only possibility is to choose the total number of electrons. Thus, we are unable to directly evaluate the conductance corresponding to the transmission peaks at $V_g^{up}$ or $V_g^{down}$, and instead we get an order-of-magnitude approximation by using the conductance of the peak close to the Fermi energy as a reference: there we find a strong conductance at low bias voltages (approx. 2 mS at $V_{bias}$<50 mV). Since this peak near $E_F$ is equally sharp but a little lower compared with the transmission peaks of our interest, and since this peak is slightly off-resonance compared with the Fermi energy, this is a rather conservative estimate. While this is a limitation of our theoretical estimation of the conductance at the relevant gate voltages, note that a series of general experimental and theoretical diffuculties in this research field hamper the accuracy of calculated conductances in every case, as discussed in more length in SI-3. Importantly, this uncertainty in the details of each particular molecule, does not affect the analysis in chemical terms we perform below.

Note that an alternative molecule showing a similar behavior is $[V(\alpha-C_3S_5)_3]^{2-}$, a similar complex where the atom contacting the Au(111) electrodes is sulfur instead of oxygen and with a third dithiolate ligand coordinating the vanadium atom. For this system our calculations estimate that the gate voltage difference needed to switch the spin polarization is $\Delta(V_g) = 0.25$ V, i.e. it is much larger than that of $[VO(C_3S_4O)_2]^{2-}$. However, the better S-Au contact results in this case in a spin-polarized conductance that can be up to an order of magnitude higher compared with that obtained for $[VO(C_3S_4O)_2]^{2-}$.

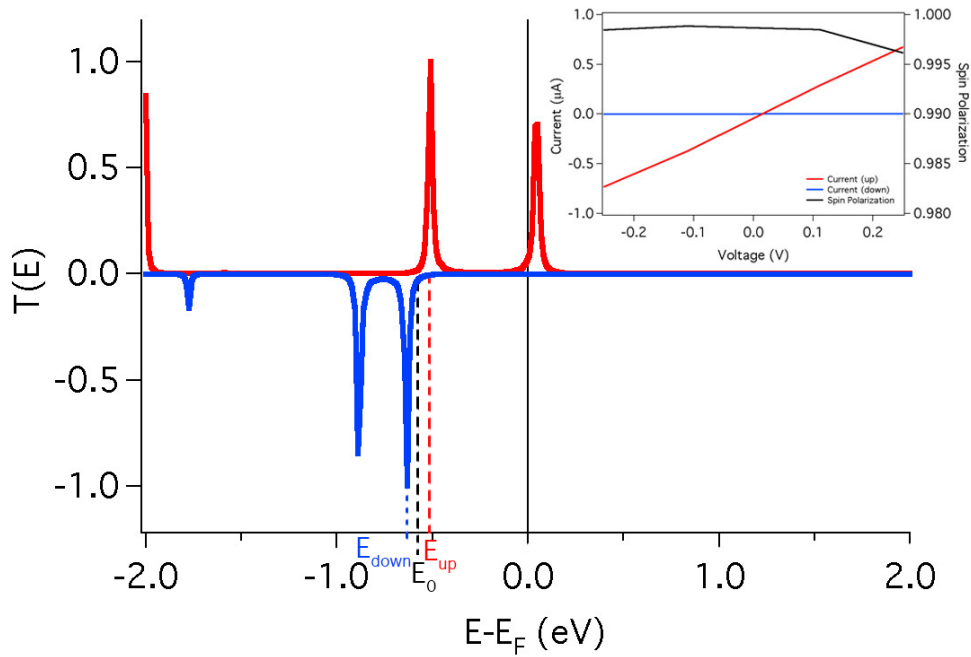

Figure 2: Transmission function calculated for [VO(C$_3$S$_4$O)$_2$]$^{2-}$ in contact to Au electrodes [ inset: Intensity-voltage plot for spin up (red) and spin-down (blue) and the resultant total spin polarization (black).]

The large spin polarization of the current can be drastically reduced by inelastic spin flip effects. In particular this spin-relaxation channel is active for minority spin transport, namely where the spin polarization of the moving electron is opposite to the molecule magnetization. The probability for this process is enhanced when the molecular state responsible for the spin transport overlaps significantly with the magnetic orbitals of the molecule, in this case with the orbitals associated to the V center. Thus we theoretically estimate its likelihood by calculating the local density of states centered on the conduction peaks, following the procedure detailed in Ref. 13. As can be seen in Figure 3, both for the spin-up and for the spin-down peaks, the magnetic orbitals barely participate in the conduction channel, meaning that in principle we should expect the molecular spin to be stable even when filtering minority carriers.

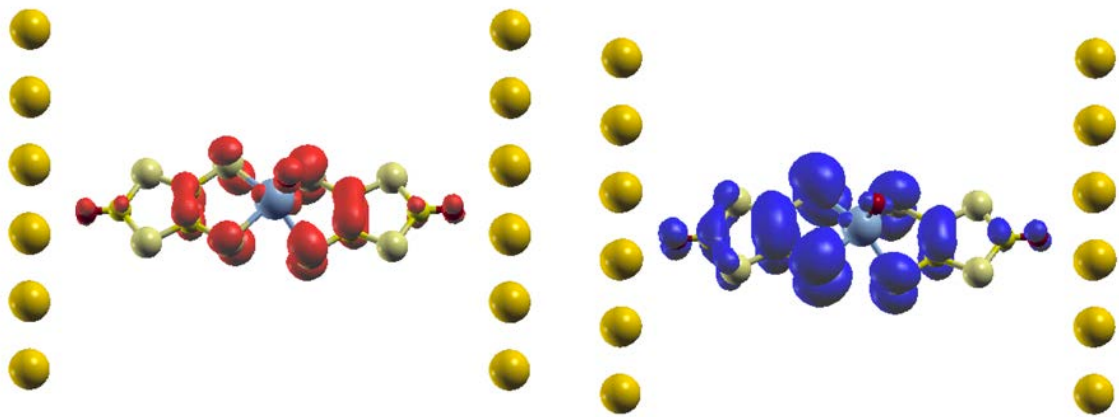

Figure 3: Local density of states associated to the two conduction peaks (left: $V_g^{up}$=-0.51V, right: $V_g^{down}$=-0.63V) involved in the dual spin-filter process. The spin density is almost exclusively up (red) for $V_g^{up}$ and down (blue) for $V_g^{down}$.

In summary, we can now propose $[VO(C_3S_4O)_2]^{2-}$ and $[V(\alpha-C_3S_5)_3]^{2-}$ to be interesting molecules for creating dual spin filter molecular spintronics devices. We can expect this behavior in a wider spectrum of mononuclear magnetic complexes, as long as they present two strong and sharp conduction channels of opposite spin polarities in a limited energy range and in presence of a fixed polarizing magnetic field. In chemical terms, molecules need to establish only a moderate hybridization with the electrodes (sharp channels), and at the same time they need to be intrinsically strongly conductive, i.e. they need to be highly delocalized over the molecule. In the present example this is achieved by choosing a small molecule, $[VO(C_3S_4O)_2]^{2-}$, with a good aromatic channel but a weak contact with Au(111), due to the relatively weak O-Au bonds rather than the stronger and more common S-Au bonding. Additionally, the molecular spin state needs to survive to inelastic spin-flip in the conduction regime, meaning that they need to reside over different orbitals than those associated to the magnetic center. Since these requisites are not particularly exotic, one can appreciated that $[VO(C_3S_4O)_2]^{2-}$-Au(111) should not be a rarity but rather a member of a wide set of molecular complexes that can perform dual spin filtering via a minimal switching gate voltage.

The authors declare no competing financial interests.

ACKNOWLEDGMENT

The research reported here was supported by the Spanish MINECO (Grants MAT 2014-56143-R and CTQ 2014-52758-P co-financed by FEDER, and Excellence Unit María de Maeztu MDM-2015-0538), the European Union (ERC-CoG DECRESIM 647301 and COST-MOLSPIN-CA15128 Molecular Spintronics Project), and the Generalitat Valenciana (Prometeo Program of Excellence). A. Gaita-Ariño thanks the Spanish MINECO for a Ramón y Cajal Fellowship. S. Cardona-Serra acknowledges the Generalitat Valenciana for a VAL i+D postdoctoral contract. S. Sanvito acknowledges the Quest project funded by European Research Council. E. Navarro-Moratalla acknowledges financial support from Generalitat Valenciana (subvenciones a la excelencia científica Juniors Investigadores de la Generalitat Valenciana). We thank the Trinity Centre for High Performance Computing (TCHPC) for providing the computational resources.

REFERENCES

1  Baibich, M. N.; Broto, J. M.; Fert, A.; Van Dau, F. N.; Petroff, F.; Etienne, P.; Creuzet, G.; Friederich, A.; Chazelas, J. *Physical Review Letters* **1988**, 61 (21), 2472.

2   Binasch, G.; Grünberg, P.; Saurenbach, F.; Zinn, W. *Physical Review B* **1989**, *39* (7), 4828.

3   Pan, C.; Chang, S.-C.; Naeemi, A. IEEE, **2016**; 56–58.

4   Sanvito, S. *Chem. Soc. Rev.*, **2011**, *40*, 3336.

5   Dediu, V. A.; Hueso, L. E.; Bergenti, I.; Taliani, C. *Nat. Mater.* **2009**, *8*, 707.

6   Bogani, L.; Wernsdorfer, W. *Nat. Mater.* **2008**, *7*, 179.

7   Awschalom, D. D.; Flatté, M. E. *Nat Phys* **2007**, *3* (3), 153

8   Godfrin, C.; Ferhat, A.; Ballou, R.; Klyatskaya, S.; Ruben, M.; Wernsdorfer, W.; Balestro, F. *Physical Review Letters* **2017**, *119* (18), 187702.

9   Droghetti, A.; Thielen, P.; Rungger, I.; Haag, N.; Großmann, N.; Stöckl, J.; Stadtmüller, B.; Aeschlimann, M.; Sanvito, S.; Cinchetti, M. *Nature Communications* **2016**, *7*, 12668.

10  Baadji, N.; Piacenza, M.; Tugsuz, T.; Della Sala, F.; Maruccio, G.; Sanvito, S. *Nat. Mater.* 2009, **8**, 813.

11  Ozaki, T.; Nishio, K.; Weng, H.; Kino, H. *Phys Rev B* **2010**, *81* (7), 1018

12  Pemmaraju, C. D.; Rungger, I.; Sanvito, S. *Phys Rev B* **2009**, *80* (10), 104422.

13  Cardona-Serra, S.; Gaita-Ariño, A.; Stamenova, M.; Sanvito, S. *J. Phys. Chem. Lett.* **2017**, 3056.

14  Wan, H.; Zhou, B.; Chen, X.; Sun, C. Q.; Zhou, G. *J. Phys. Chem. C* **2012**, *116* (3), 2570.

15  Wu, Q.-H.; Zhao, P.; Liu, D.-S.; Li, S.-J.; Chen, G. *Organic Electronics* **2014**, *15* (12), 3615.

16  Zhou, Y.-H.; Zeng, J.; Tang, L.-M.; Chen, K.-Q.; Hu, W. P. *Organic Electronics* **2013**, *14* (11), 2940.

17  (a) Zadrozny, J. M.; Niklas, J.; Poluektov, O. G.; Freedman, D. E. *ACS Central Science* **2015**, *1* (9), 488. (b) Yu, C.-J.; Graham, M. J.; Zadrozny, J. M.; Niklas, J.; Krzyaniak, M. D.; Wasielewski, M. R.; Poluektov, O. G.; Freedman, D. E. *J. Am. Chem. Soc.* **2016**, *138* (44), 14678.

# Supporting Information for:

# Proposal for a dual spin filter based on [VO(C$_3$S$_4$O)$_2$]$^{2-}$

*S. Cardona-Serra*[*,1,2], *A. Gaita-Ariño*[1], *E. Navarro-Moratalla*[1] *and S. Sanvito*[2].

[1] Instituto de Ciencia Molecular, Universitat de València, España.

[2] School of Physics, AMBER and CRANN Institute, Trinity College, Dublin 2, Ireland.

Contents:

1: Theoretical Methodology

2: Determination of the density of states and I-V curves for the bis- and tris- dithiolate complexes.

3: On the molecular trapping between two electrodes.

4: Previous molecular spin-filtering estimates

5: References

1: Theoretical Methodology

Transport calculations are performed using the SMEAGOL code[1] that interfaces the non-equilibrium Green's function (NEGF) approach to electron transport with the density functional theory (DFT) package SIESTA[2]. All the structural relaxation calculation were performed using the original SIESTA code.

In our simulations the transport junction is constructed by placing the molecule between two Au (111)-oriented surfaces with 5×5 cross section except the complex $[V(C_8S_8)_3]_{2-}$, which was sandwiched between two 7x7 Au (111) surface due to its larger size. This mimics a standard transport break-junction experiment with the most used gold surface orientation. The choice of a gold electrode arises from stability of the sulfur-gold bond that ensures the best attachment between the molecule and the surfaces. The initial NM-surface (NM= S, O) distance was set to 2.0 Å with the NM atom located at the 'hollow site', the most stable absorption position discussed in literature.[3] Thus, the entire structure is then relaxed until the maximum atomic forces are less than 0.01 eV/Å. A real space grid with an equivalent plane wave cutoff of 200 Ry (enough to assure convergence) has been used to calculate the various matrix elements. Finally the electronic temperature of the calculation is set to 0.1 K to mimic the low-temperature conditions needed in a typical break-junction experiment.

The exchange-correlation potential is described by the Ceperley-Alder local density approximation (LDA)[4] as implemented for using the ASIC approach (vide infra). Self-interaction error is one of the main problems when using local-density functionals for the study of the interfaces between inorganic leads and molecules. This error leads to an over-delocalization of the electronic density producing the HOMO orbitals to be artificially raised in energy. The LUMO orbitals could also be found at lower energy values that the commonly expected. In order to avoid this problem we have used the Atomic

Self Interaction Correction (ASIC) method.[5] This methodology sets adequately the position of the HOMO orbital and improves the energy level alignment obtaining values of conductance in with a better correspondence with experiments.[6]

The Au-valence electrons are represented over a numerical s-only single-z basis set that has been previously demonstrated to offer a good description of the energy region around the Fermi level.[7] In contrast, for the other atoms (S, C, V and O) we use a full-valence double-z basis set with polarization (basis size was increased until convergence). Norm-conserving Troullier-Martins pseudopotentials[8] (previously adapted to ASIC methodology) are employed to describe the core-electrons in all the cases. Considering the fact that all the molecules studied here are dianions, we have shifted the energy levels obtained by adding (removing) a specific number of charges depending of the molecular charge. The limitation of this approach is that these charges are only added for the calculation of the electrostatic potential but do not enter in the self-consistent charge density, thus they will only interact indirectly with the charges of the scattering region via the electrostatic potential. (See SMEAGOL manual for more details about calculations on charged systems).

Finally, the spin-dependent current, $I_s$, flowing through the junction is calculated from the Landauer-Büttiker formula[9],

$$I_\sigma(V) = \frac{e}{h}\int_{-\infty}^{+\infty} T_\sigma(E,V)[f_L(E-\mu_L) - f_R(E-\mu_R)]dE$$

where the total current $I_{tot}$ is the sum of both the spin-polarized components, $I_s$, s = spin up /spin down. Here $T_\sigma(E,V)$ is the transmission coefficient[1] and $f_{L/R}$ are the Fermi functions associated to the two electrodes chemical potentials, $\mu_{L/R} = \mu_o \pm V/2$, where $\mu_o$ is the electrodes common Fermi level.

In our two-spin-fluid approximation (there is no spin-flip mechanism) majority and minority spins carry two separate spin currents and the resultant current spin polarization, SP, is calculated as

$$SP = \frac{I_{up} - I_{down}}{I_{up} + I_{down}}$$

Note that although this approach is widely used to perform theoretical predictions within the spintronic community,[10] it has some limitations inherent to the basics of DFT methodology as pointed out by Lozano et al.[11] One may note that the theoretical spin asymmetry corresponds to what is expected from an open-shell DFT calculation in a magnetic molecule. Thus, within the mono-determinant approximation of DFT only a concrete $M_S$ state is calculated. In literature this has been sometimes confused with the obtainment of a super-efficient spin filter. This is a different scenario compared with our case, where the low temperature guarantees the spin orientation and it is experimentally fixed by an external polarizing magnetic field.

2: Determination of the density of states and I-V curves for the bis- and tris- dithiolate complexes.

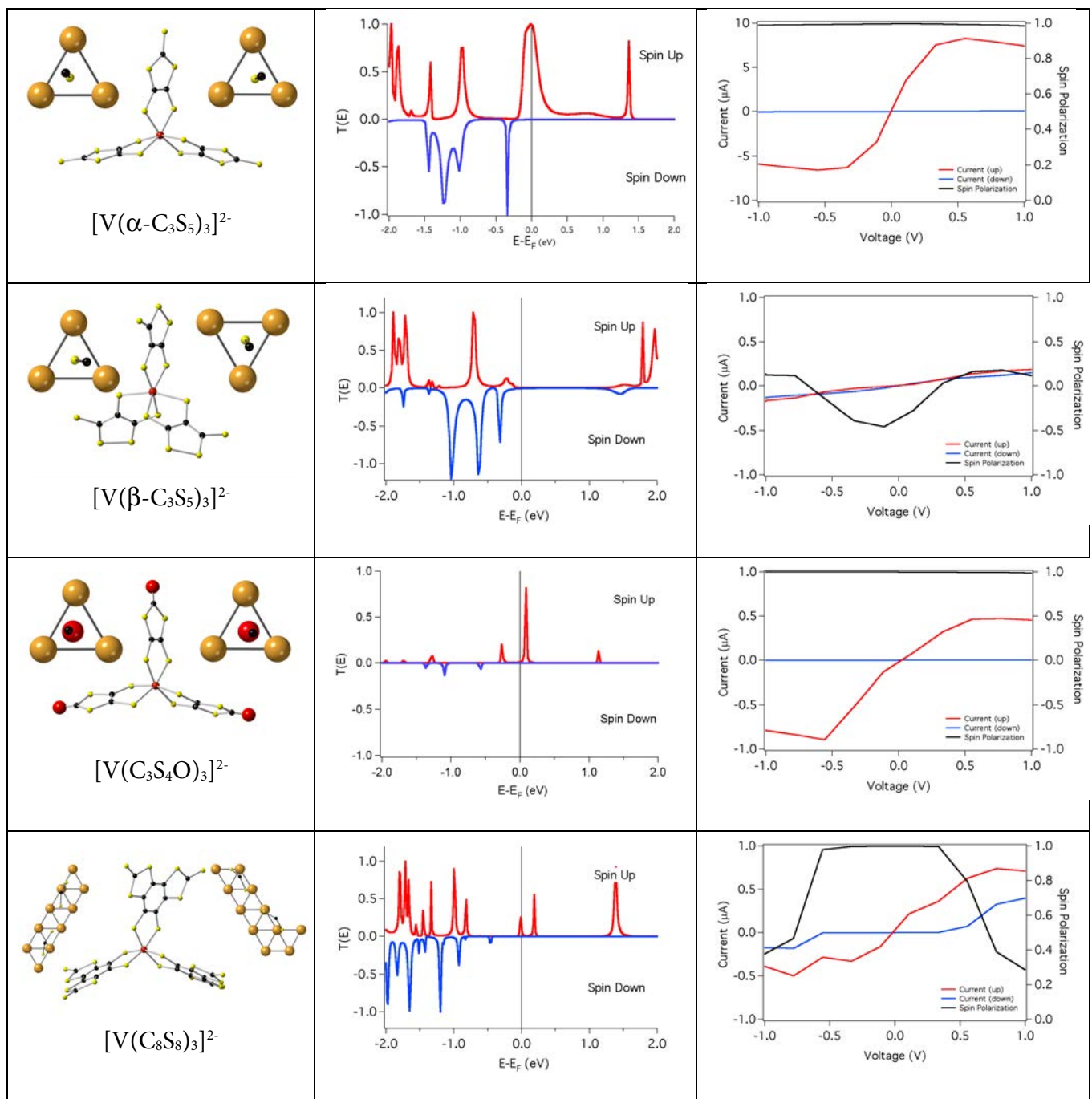

Figure S.1: Left: Chemical structure of the tris-dithiolate anions. Inset: Lateral view of the NM-Au attachment. Centre: Calculated Density of states spectra. Right: Current-Voltage curves.

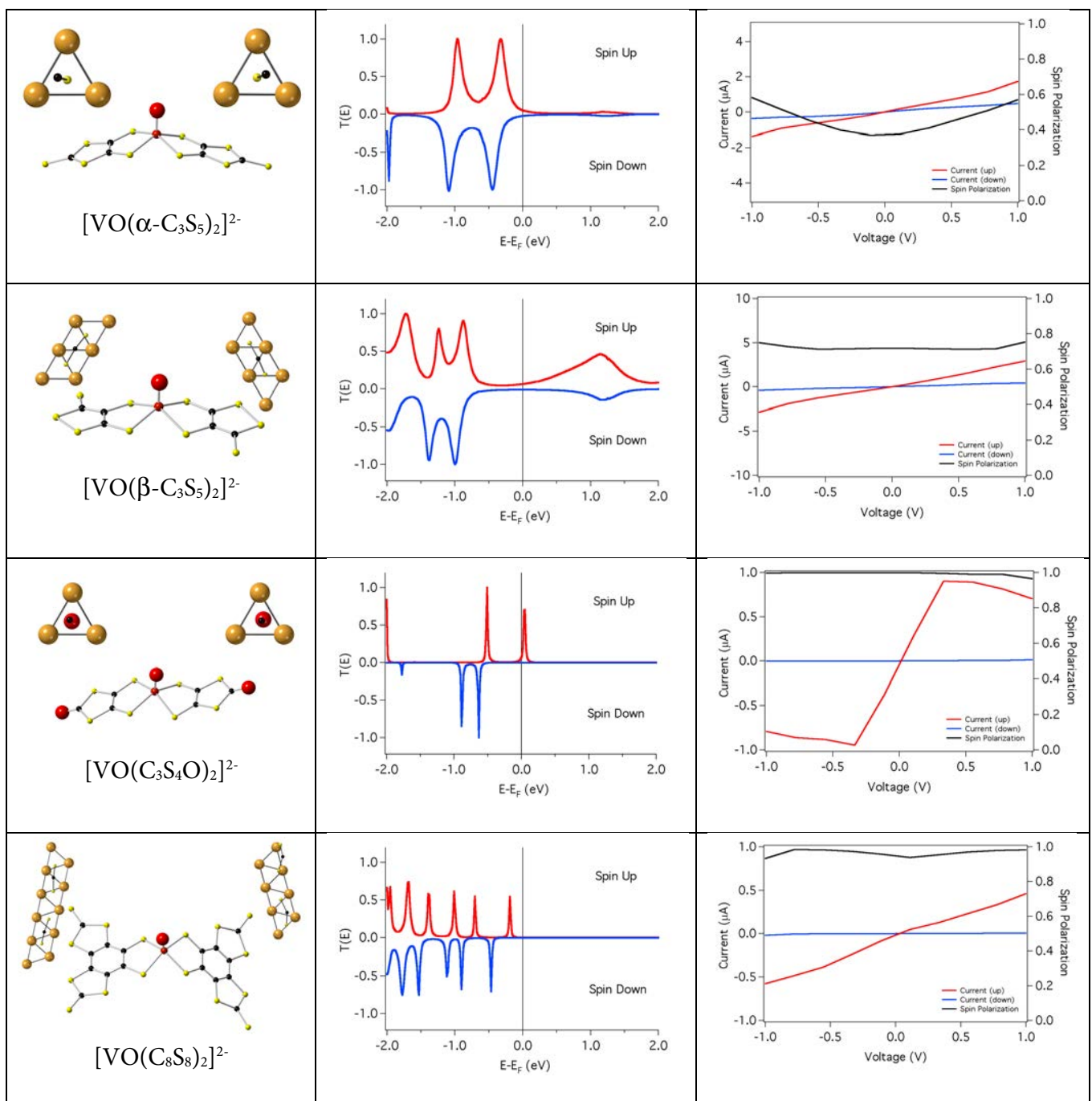

Figure S.2: Left: Chemical structure of the tris-dithiolate anions. Inset: Lateral view of the NM-Au attachment. Centre: Calculated Density of states spectra. Right: Current-Voltage curves.

Let us focus first on the first series, the vanadium tris-dithiolates. In terms of the gold – nonmetal bond, there are significant differences between, on the one side, the cases of $[V(\alpha-C_3S_5)_3]^{2-}$ and $[V(\beta-C_3S_5)_3]^{2-}$ (Fig S1, Row 1, 2), which are directly linked by a single sulfur atom located in the energetically most favorable hollow space and, on the other side, the cases of $[V(C_3S_4O)_3]^{2-}$ (Fig S1, Row 3), which is linked by a single oxygen in the hollow space and $[V(C_8S_8)_3]^{2-}$ (Fig S1, Row 4), where four different sulfur atoms need to be accommodated in the most preferable locations while respecting the chemical structure of the molecule. In the two former cases the strong Au-S bonding allows a higher hybridization between molecule and electrodes, thus resulting in significantly wider and higher peaks in the density of states. In particular, a good compromise between selectivity and conductance is found for $[V(\alpha-C_3S_5)_3]^{2-}$ and can not be observed for $[V(\beta-C_3S_5)_3]^{2-}$. Contrarily, in $[V(C_3S_4O)_3]^{2-}$, the presence of relatively weak Au-O bonds diminishes the hybridization; as a result, the obtained peaks are considerably sharp but sparse, making them useless as dual spin filters. In the latter case, $[V(C_8S_8)_3]^{2-}$, the optimal attachment cannot be achieved without jeopardizing the integrity of the molecule, since the most preferable positions are far enough to be reached by all the S atoms. Thus provokes a compromise situation between the bonding stabilization (Au-S) versus the stability of the ligand. This results in an effective weak attachment and thus in sharp peaks in the DOS. Unfortunately, this complex offers a dense spectrum with many sharp but close-lying peaks impeding the selection of this complex as a valid candidate.

Focusing on the second series, the vanadyl bis-dithiolates, we can again observe similar trends compared with trisdithiolates. The two first complexes, $[VO(\alpha-C_3S_5)_2]^{2-}$ and $[VO(\beta-C_3S_5)_2]^{2-}$ (Fig S2, Rows 1, 2), present the widest peaks among all studied cases, thus we deduce that the hybridization is very strong in these cases. The origin of this high interaction is likely related with the absence of a third ligand where the charge can be delocalized. This absence might provoke an increase in the electronic density located in the Au-S bond and thus a major and stronger coupling between molecule and electrode. The two latter complexes [VO($C_3S_4O$)$_2$]$^{2-}$ (Fig S2, Row 3) and [VO($C_8S_8$)$_2$]$^{2-}$ (Fig S2, Row 4), present sharper DOS peaks which make them promising in principle, but again, as happened with the previous series, the complex [VO($C_8S_8$)$_2$]$^{2-}$, with larger ligands, possesses a richer and more complex DOS spectrum. This goes against its potential as a selective spin filter.

3: On the molecular trapping between two electrodes

One of the main experimental limitations in single-molecule spintronics is the poor reproducibility in the attachment of the molecules to the electrodes. Small differences in the electrode surface can have critical consequences in the electric (and magnetic) response of the molecular device, and break junction setups are inherently prone to at least minor irreproducibilities. Furthermore, it is not straightforward to guarantee the 'cleanliness' in the molecular device: counterions, solvent molecules or even a nearby magnetic molecule can modify the behavior in critical ways. Finally, even with an isolated molecule between two reproducible surfaces at a given distance, there is often more than one stable molecule-electrode configuration. In sum, it is understandable that even cutting-edge works in the field need to refer to "sample A", "sample B", "sample C" to discuss the different experiments where uncontrolled parameters resulted in different electric and magnetic responses.[12]

This experimental problem is compounded with calculations, since it is far from trivial to determine the truly fundamental configuration even for an isolated molecule between two Au(111) plates. As a consequence, we need to be very careful in the interpretations of our theoretical predictions. It would be the scope of a more comprehensive study to obtain the relaxed structures starting from a wider set of electrode-molecule-electrode configurations, and even this would not solve the experimental problem of reproducibility.

In the case of the present work, the details of our calculations will change with different structural arrangements, but the insights we acquired regarding molecular design are robust: in order to present dual spin filtering behavior, one needs a relatively simple transmission spectrum, with few, sharp, strong transmission channels (these are necessary but not sufficient conditions). In chemical terms, this can be achieved by choosing a magnetic molecule that

-is mononuclear (and thus presents less transmission channels than a polynuclear complex)

-has a weak bonding to the electrodes (and thus presents sharp channels) and

-is small and/or contains a good pi cloud (and thus has strong conductance).

In addition, the spin needs to be localized in a magnetic orbital that is either orthogonal or at a different energy compared with the conduction orbitals.

These criteria, while offering no guarantee, may guide the experimental efforts towards molecules that at least are good candidates to observe the desired behavior.

## 4: Previous molecular spin-filtering estimates:

The idea of using molecules for spin-filtering is not new in itself. However, previous theoretical explorations of spin-filtering effects using unimolecular devices have been severely lacking. In a first example, the molecular structure was highly unrealistic from the chemical point of view, and the external stimulus for the switching was also ill-defined. This is the situation in the case of "Switching, Dual Spin-Filtering Effects, and Negative Differential Resistance in a Carbon-Based Molecular Device",[13] where the proposed molecular device is based on the connection of two 4-carbon wide graphene nanoribbons, which act as leads, to a 1,4-bis(pentatetraene)benzene, which supposedly acts as switch. Moreover, the switching itself is also purely theoretical, since it requires a microscopic 90 degree rotation of the 1,4 bis(pentatetraene)benzene with respect to the leads, with no clear indication of how to achieve such a controlled rotation.

A similar situation is found for "Giant magnetoresistance effect and spin filters in phthalocyanine based molecular devices",[14], where the key to the behavior is the chemical fusion between a phthalocyanine molecule with two 8-carbon wide zig-zag graphene nanoribbons. In this case, however, no good dual spin filtering is observed, presumably due to the excellent contact between molecule and electrodes, which translates into wide, overlapping transmission peaks.

A more extreme case is found for "Effects of the magnetic anchoring groups on spin-dependent transport properties of Ni(dmit)$_2$ device",[15] where dual spin filtering is predicted for a molecule that is theoretically derived from Ni(dmit)$_2$ by substituting a sulphur atom by a Ni atom in one end of the molecule, and the equivalent sulphur atom in the opposite end by a Mn atom, and having the Ni and Mn atoms act as bridges to the Au electrodes. While studies of this kind may have theoretical interest, they hold no potential for experimental realization.

In two further studies, spin filtering was only found in a single direction, a comparatively trivial achievement, with no possibility for dual filtering. This is the case of "Rectifying, giant magnetoresistance, spin-filtering, newgative [sic] differential resistance, and switching effects in single molecule magnet Mn(dmit)$_2$-based molecular device with grapheme nanoribbon electrodes".[16]

# 5: References

(1) (a) Rocha, A. R.; Garcia-Suarez, V.; Bailey, S. W.; Lambert, C. J.; Ferrer, J.; Sanvito. S. Towards Molecular Spintronics. *Nat. Mater.* **2005**, *4*, 335-339. (b) Rungger, I.; Sanvito, S. Algorithm for the Construction of Self-Energies for Electronic Transport Calculations Based on Singularity Elimination and Singular Value Decomposition. *Phys. Rev. B* **2008**, *78*, 035407.

(2) Soler, M.; Artacho, E.; Gale, J. D.; Garcia, A.; Junquera, J.; Ordejon, P.; Sánchez Portal, D. The SIESTA Method for Ab-Initio Order-N Materials Simulation. *J. Phys.: Condens. Matter*, **2002**, *14*, 2745-2779.

(3) Sellers, H.; Ulman, A.; Shnidman, Y.; Eilers, J. E. Structure and Binding of Alkanethiolates on Gold and Silver Surfaces: Implications for Self-Assembled Monolayers. *J. Am. Soc. Rev.* **1993**, *115*, 9389-9401.

(4) Ceperley, D.M.; Alder, B. J. Ground State of the Electron Gas by a Stochastic Method. *Phys. Rev. Lett.* **1980**, *45*, 566-569.

(5) (a) Pemmaraju, C. D.; Archer, T.; Sánchez-Portal, D.; Sanvito, S. Atomic-Orbital-Based Approximate Self-Interaction Correction Scheme for Molecules. *Phys. Rev. B*, **2007**, *75*, 045101. (b) Filippetti, A.; Pemmaraju, C. D.; Sanvito, S.; Delugas, P.; Puggioni, D.; Fiorentini, V. Variational Pseudo-Self-Interaction-Corrected Density Functional Approach to the Ab-Initio Description of Correlated Solids and Molecules. *Phys. Rev. B,* **2011**, *84*, 195127-195149. (c) Toher, C.; Filippetti, A.; Sanvito, S.; Burke, K. Self-Interaction Errors in Density Functional Calculations of Electronic Transport. *Phys. Rev. Lett.*, **2005**, *95*, 146402-146406.

(6) (a) Pontes, R. B.; Rocha, A. R.; Sanvito, S.; Fazzio, A.; Da Silva, J. R. Ab-Initio Calculations of Structural Evolution and Conductance of Benzene-1,4-Dithiol on Gold Leads. *ACS Nano*, **2011**, *5*, 795-804. (b) French, W. R.; Iacovella, C. R.; Rungger, I.; Souza, A. M.; Sanvito, S.; Cummings, P. T. Structural Origins of Conductance Fluctuations in Gold-Thiolate Molecular Transport Junctions. *J. Phys. Chem. Lett.* **2013**, *4*, 887-891. (c) French, W. R.; Iacovella, C. R.; Rungger, I.; Souza, A. M.; Sanvito, S.; Cummings, P. T. Atomistic Simulations of Highly Conductive Molecular Transport Junctions Under Realistic Conditions. *Nanoscale.* **2013**, *5*, 3654-3663.

(7) Toher, C.; Sanvito, S. Effects of Self-Interaction Corrections on the Transport Properties of Phenyl-Based Molecular Junctions. *Phys. Rev. B,* **2008**, *77*, 155402.

(8) Troullier, N.; Martins, J. L. Efficient Pseudopotentials for Plane-Wave Calculations. *Phys. Rev. B* **1991**, *43*, 1993-2006.

(9) Büttiker, M.; Imry, Y.; Landauer, R.; Pinhas, S. Generalized Many-Channel Conductance Formula with Applications to Small Rings. *Phys. Rev. B,* **1985**, *31*, 6207-6215.

(10) Sanvito, S. Ab-initio Methods for Spin-Transport at the Nanoscale Level. In Handbook of Computational Nanotechnology Vol 5. American Scientific Pubishers, California, USA, 2004. Also available at arXiv:cond-mat/0503445 and references herein.

(11) Kepenekian, M.; Gauyacq, J.P.; Lorente, N. Difficulties in the Ab-Initio Description of Electron Transport Through Spin Filters. *J. Phys.: Condens. Matter*, **2014**, *26,* 104203

(12) Godfrin, C.; Thiele, S.; Ferhat, A.; Klyatskaya, S.; Ruben, M.; Wernsdorfer, W.; Balestro, F. *ACS Nano,* **2017**, *11* (4), 3984. (See supporting information).

(13) Wan, H.; Zhou, B.; Chen, X.; Sun, C. Q.; Zhou, G. *J. Phys. Chem. C* **2012**, *116* (3), 2570

(14) Zhou, Y.H.; Zeng, J.; Tang, L.M.; Chen, K.Q.; Hu, W.P. *Organic Electronics* **2013**, *14* (11), 2940-2947.

(15) Yan, S.; Long, M.; Zhang, X.; He, J.; Xu, H.; Chen, K. *Chemical Physics Letters* **2014**, *608*, 28-34]

(16) Wu, Q. H.; Zhao, P.; Liu, D.-S.; Li, S.-J.; Chen, G. *Organic Electronics* **2014**, *15* (12), 3615